\documentclass[10pt]{IEEEtran}
\makeatletter
\def\ps@headings{%
\def\@oddhead{\mbox{}\scriptsize\rightmark \hfil \thepage}%
\def\@evenhead{\scriptsize\thepage \hfil \leftmark\mbox{}}%
\def\@oddfoot{}%
\def\@evenfoot{}}
\makeatother 
\usepackage{amssymb,amsmath}
\usepackage{color,soul}
\usepackage{amsfonts}
\usepackage{array}
\usepackage{algorithm,algorithmic}
\usepackage[nodisplayskipstretch]{setspace}
\usepackage{mathrsfs}
\usepackage{graphicx}
\usepackage{multirow}
\usepackage{epstopdf}
\usepackage{epsfig}
\usepackage{stmaryrd}
\usepackage{multirow,url}
\usepackage{color,}
\usepackage{amsthm}
\usepackage{bm}
\usepackage{pgf,tikz}
\usepackage{subfigure,cite}

\usepackage{fancyhdr,ifthen}

\DeclareMathOperator{\Exp}{\mathbb{E}}

\newcounter{section:outage-analysis}
\begin{document}

\title{Joint Time-Domain Resource Partitioning, Rate Allocation, and Video Quality Adaptation in Heterogeneous Cellular Networks}

\author{Antonios Argyriou,~\IEEEmembership{Member,~IEEE}, Dimitrios Kosmanos, Leandros Tassiulas,~\IEEEmembership{Fellow,~IEEE}\thanks{Antonios Argyriou and D. Kosmanos are with the Department of Electrical and Computer Engineering, University of Thessaly, Greece. L. Tassiulas is with Yale University, USA.}}

\graphicspath{{figures/}}

\maketitle

\begin{abstract}
Heterogenous cellular networks (HCN) introduce small cells within the transmission range of a macrocell. For the efficient operation of HCNs it is essential that the high power macrocell shuts off its transmissions for an appropriate amount of time in order for the low power small cells to transmit. This is a mechanism that allows  time-domain resource partitioning (TDRP) and is critical to be optimized for maximizing the throughput of the complete HCN. In this paper, we investigate video communication in HCNs when TDRP is employed. After defining a detailed system model for video streaming in such a HCN, we consider the problem of maximizing the experienced video quality at all the users, by jointly optimizing the TDRP for the HCN, the rate allocated to each specific user, and the selected video quality transmitted to a user. The NP-hard problem is solved with a primal-dual approximation algorithm that decomposes the problem into simpler subproblems, making them amenable to fast well-known solution algorithms. Consequently, the calculated solution can be enforced in the time scale of real-life video streaming sessions. This last observation motivates the enhancement of the proposed framework to support video delivery with dynamic adaptive streaming over HTTP (DASH). Our extensive simulation results demonstrate clearly the need for our holistic approach for improving the video quality and playback performance of the video streaming users in HCNs.
\end{abstract}


\section{Introduction}
\begin{figure}[t]
\centering
\includegraphics[keepaspectratio,width = 0.95\linewidth]{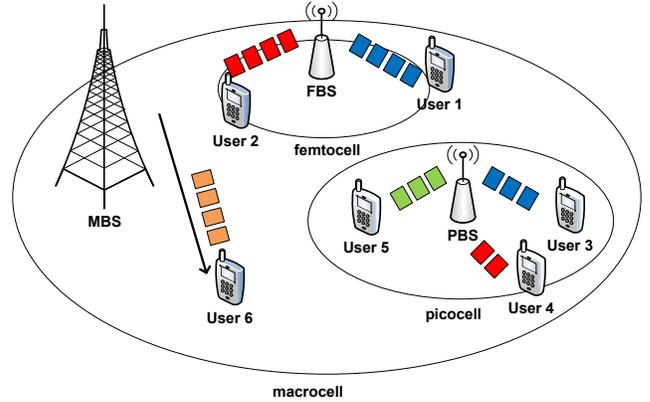}%
\caption{The considered HCN consists of single macro and several pico and femto BSs. Each BS streams videos to a subset of the associated users. }
\label{fig:system-model}
\end{figure}

\IEEEPARstart{N}{early} 50\% of the traffic in cellular networks today is video~\cite{ericsson-2014}. Mounting evidence suggests that video will keep increasing its share of the cellular traffic at an even faster pace~\cite{ericsson-2014}. The reason behind this phenomenon is the explosive demand for high quality video streaming from mobile devices (e.g., tablets, smart-phones). The challenge for mobile network operators (MNOs) is to offer higher data rates that can keep up with this demand for high quality video. Heterogenous cellular networks (HCNs), illustrated in Fig.~\ref{fig:system-model}, are envisioned to be one of the solutions to this problem. HCNs introduce low power base stations (BS) like pico BS (PBS) and femto BS (FBS), that form around them picocells and femtocells respectively. Lower transmission power from these small cells reduces the transmission range and allows improved spatial reuse. Hence, the first novel feature of HCNs is the \textit{higher wireless capacity} they offer to the complete macrocell. The second novel feature of HCNs is that they can \textit{lower the use of the backhaul capacity} by employing caching at the small cell BSs~\cite{femtocaching,poularakis-infocom}. Caching enables local  access of frequently requested videos and this means lower utilization of the backhaul links between a BS and the video server. These two features of HCNs constitute them a central component of the envisioned 5G cellular network architecture.

\begin{figure}[t]%
\centering
\includegraphics[keepaspectratio,width = 0.95\linewidth]{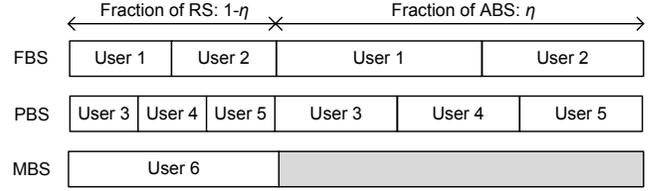} 
\caption{Modeling the global time-domain resource partitioning, and local rate allocation problem in a topology with two PBSs and a MBS.}
\label{fig:problem-model}
\end{figure}

Research for video distribution in HCNs has focused primarily on caching, with the objective to reduce the startup playback delay of the video for each user~\cite{femtocaching,poularakis-infocom}, or lower the costs for the operator~\cite{poularakis-infocom}. In this paper we are concerned with the first novel feature of HCNs which is the higher wireless capacity. We focus on this topic since HCNs introduce a new way for sharing the wireless resources. With the time domain resource partitioning (TDRP) mechanism~\cite{LTE12}, the MBS shuts off its transmissions for a fraction $\eta$ of the available resources during which the small cells can achieve a higher data rate (Fig.~\ref{fig:problem-model}). During the fraction $1-\eta$, there is \textit{intra-cell} interference since the MBS transmits simultaneously with the small cells. This technique was recently standardized through the concept of almost blank subframes (ABS) and regular subframes (RS) in 3GPP LTE-A under the more general name of enhanced inter-cell interference coordination (eICIC)~\cite{LTE12}. One important detail is that the LTE-A standard currently allows the dynamic adaptation of $\eta$ but it does not specify how it should be configured. Given the increasing number of video streaming users in cellular networks, and the necessity of TDRP, it is of outmost importance to perform optimally both the configuration of $\eta$ and the allocation of the wireless resources in an HCN. Hence, the specific questions that should be answered in this case are: What is the optimal $\eta$ when we have video traffic? What is the best video quality that each user should receive? What happens when a subset of the users receive video? Currently there are no definite answers to these pressing questions.

\textbf{Related work.} TDRP for HCNs is a topic investigated only recently because ABS/RS were also very recently standardized in LTE-A. The authors in~\cite{monogioudis14} derived the optimal fraction from the available ABS and RS resources that each user should be allocated (a representative rate allocation is illustrated in Fig.~\ref{fig:problem-model}) under a proportionally fair rate allocation (PFRA) metric. The authors of that work assumed a constant fraction of ABS $\eta$ that is configured by the HCN operator. In another recent work reported in~\cite{andrews14b}, the authors investigated the joint optimization of TDRP and user association (for traffic offloading) but with an assumption for equal rate allocation to the associated users. To the best of our knowledge there is no work that addresses TDRP in HCNs for video distribution and streaming. As we already discussed, much of the early research work for video streaming HCNs has focused on caching~\cite{femtocaching,poularakis-infocom}, or exploiting particular features like the density of the small cells~\cite{argyriou:jnl_2015_spic}. However, these works assumed the availability of a constant fraction of the resources for the MBS and the small cells that is effectively translated to a constant $\eta$. On the other hand, multi-user rate allocation for video streaming has been a topic thoroughly investigated the last few years for specific types of wireless networks. The works were primarily motivated from the network utility maximization (NUM) framework \cite{Shakkottai07}. From the category of works that were based on NUM, the ones that are more related to this paper focused on cellular networks and considered more details of the physical layer (PHY). For example the authors in~\cite{Chiang13} investigated scheduling and resource allocation for a downlink LTE system that employs discrete decisions for optimizing the selected video streaming quality. The same problem, but for scalable encoded video, was considered in~\cite{Ahmedin13}. Another class of works focused on optimizing rate/resource allocation with the objective to improve the playback performance of video clients~\cite{andrews-video-capacity,veciana14}. In the last works the authors take into consideration recent standardization developments in streaming and in particular dynamic adaptive streaming over HTTP (DASH). However, these works do not target HCNs and assume access to fixed capacity resources.

\textbf{Contributions.} In this paper, we present contributions on three fronts. First, we present a comprehensive \textit{Joint TDRP, Rate Allocation, and Video Quality Selection (JTRAVQS)} optimization framework for video streaming in HCNs. The framework identifies the optimal TDRP $\eta$, the rate allocated to each user, and the video quality description for each user, so as to maximize the aggregate video quality in the HCN. Our framework includes additional system-level parameters like the fraction of users that receive video. The NP-hard problem is solved with a primal-dual approximation algorithm that provides an asymptotically optimal solution. Our solution approach decomposes the problem into simpler subproblems, making them amenable to fast well-known solution algorithms. Second, we propose enhancements to the basic optimization framework that allow it to support DASH-based video streaming. Additional system parameters like the buffer contents of individual users, time-dependent user population, and channel capacity with TCP, are taken into account to optimize the the playback performance~\cite{Stoica11}. Third, we present a thorough performance evaluation our main framework against: a) A video-unaware system that jointly optimizes the TDRP and rate allocation under a PFRA metric~\cite{monogioudis14}. b) From the results of our scheme and that of PFRA, we can infer the performance of a system that applies optimized rate allocation and video quality selection (RAVQS) but with a fixed TDRP~\cite{argyriou-cnf-hetnet-video}. Finally, our enhanced system for DASH, that considers the content of the playback buffer, is compared against a buffer-aware system that again uses fixed resources.

\textbf{Main Results.} Our results reveal that: i) For video streaming TDRP should be more aggressive in favor of the small cells when compared to TDRP optimization under a PFRA metric. In particular even for 4 small cells and 100 users, the optimal $\eta$ should be nearly 22.2\% higher than the optimal $\eta$ under PFRA. Video quality is improved by a factor of 50\%-70\% for this scenario. ii) Using a fixed \textit{but still optimal TDRP under a PFRA metric}, and then performing a RAVQS optimization as an afterthought, is still suboptimal. In particular for a population of 100 users and 4 small cells, the previous approach leads to an average video quality loss of 18.6\% when compared to our approach. iii) For a DASH-based system with a fixed, \textit{but again optimal TDRP under PFRA}, our optimization has more significant impact. In particular the rebuffering time/events of the clients can be reduced by more than 50\% for a static network and 60\% for a network with user churn.

\textbf{Paper Organization.} The rest of the paper is organized in the following sections. Section~\ref{section:system-model} describes in detail the system model. In Section~\ref{section:optimization} we present the formulation and the solution of the optimization problem we introduce in this paper, while its extension for DASH is presented in Section~\ref{section:optimization-dash}. Performance evaluation results are presented in Section~\ref{section:performance-evaluation}, and finally we conclude in Section~\ref{section:conclusions}.

\section{System Model and Assumptions}
\label{section:system-model}
\textbf{Network Model.} In Fig.~\ref{fig:system-model} we present the network that we study in this paper and it includes a single macrocell with a MBS, the PBSs, and the users. Each BS $j$ in the set $\mathcal{J}$ communicates with the set of associated users $\mathcal{N}_j$. We also denote with $\mathcal{F}_j\subseteq \mathcal{N}_j$ a subset of the users associated to BS $j$ that are not optimized in a video-aware fashion. This parameter allows us to investigate the possibility that a fraction $f_j=(|\mathcal{N}_j|-|\mathcal{F}_j|)/|\mathcal{N}_j|$ of the users are optimized. During the fraction $\eta$ of the ABS resources all the small cells transmit and interfere with every active user in the network. Thus, we consider \emph{resource reuse} across BSs of the same tier (PBSs in our case) which is one of the main benefits of small cells since it allows spatial reuse. The aggregate average interference power that user $i$ receives is denoted as $I_{\text{ABS},i}$. During the fraction of the non-blanked resources, or regular subframes, $1-\eta$ both the MBS and PBSs transmit and the aggregate interference power that a user receives is denoted as $I_{\text{RS},i}$.

\textbf{User Model.} Each BS $j$ transmits with unicast streaming video $i$ to the similarly denoted user. The users associate to a BS by using an signal-to-interference plus noise ratio (SINR) biasing rule~\cite{andrews14b}, i.e., a user is associated to the small cell $j$, and not the MBS, if the following is true: $SNR_{\text{PBS}_j}+Bias \geq SNR_{\text{MBS}}$. This ensures that users are offloaded to the small cells~\cite{andrews14b}. Our primary objective is a static user population similarly to the literature~\cite{femtocaching,Chiang13,andrews-video-capacity,veciana14}, since we are interested to optimize the system operation within the complete playback duration of the video. However, motivated by recent experimental results that identify slow user variations in the cell throughout the day~\cite{woo13}, we also evaluate our system for this more dynamic scenario.

\textbf{Video Streaming and Playback Model.}
Without loosing generality we assume that all the BSs are assumed to have cached the videos for all the users~\cite{femtocaching,poularakis-infocom}.\footnote{Our system can easily accommodate the case that the video stream originates from a server by considering end-to-end throughput that the network can deliver.} Now if the video representation that is transmitted to user $i$ is indexed by $r$, the average bitrate that must be sustained is
\begin{equation}
\label{eqn:Rir}
R_{ir}=\frac{S_{ir}}{T_{i}+B_{i0}} \text{ bits/sec},
\end{equation}
where $S_{ir}$ is the size of the $r$-th video representation, $T_{i}$ is the total playback time of the video and $B_{i0}$ is the duration of playable content received during startup buffering (time 0). This formulation ensures that the average probability of rebuffering events is zero~\cite{veciana14}. In the first part of our optimization in Section~\ref{section:optimization}, this is the condition we adopt since we are interested to reach a decision once for the duration of the video streaming session. However, with DASH this formula is revised in Section~\ref{section:optimization-dash}.

\textbf{Channel \& PHY Model.} Nodes use a single omni-directional half-duplex antenna. The channel from the $j$-th BS to the $i$-th user is denoted as $h_{j,i}$. The fading coefficients are independent and $h_{j,i}$$\sim$$\mathcal{CN}(PL_{i,j},1)$, i.e., they are complex Gaussian random variables with unit variance and mean equal to $PL_{i,j}$ that depends on path loss and shadowing effects according to the LTE channel model~\cite{LTE12}. All the channels are considered to be block-fading Rayleigh and quasi-stationary, that is they remain constant for the coherence period of the channel that is equal to the transmission length of the complete PHY block. Additive white Gaussian noise (AWGN) is assumed at every receiver with variance $\sigma^2$. The transmission power that the PBS and MBS use is $P_\text{PBS}$, and $P_\text{MBS}$ respectively.

\textbf{MCS \& CQI.} A modulation and coding scheme (MCS) with $m$ bits/symbol is used by each BS while its value is determined by each PBS independently and optimally as we will later explain. The set of available MCSs is $\mathcal{M}=\{1,...,7\}$, i.e., we assume that the most spectral efficient quadrature amplitude modulation (QAM) MCS is 128-QAM. We also assume that users provide only \emph{average channel quality indicator} (CQI) feedback to the BSs.

\subsection{Video Quality Model}
Modeling the Quality-of-Experience (QoE) of users in video streaming applications is not easy. QoE is affected both by the video signal quality and delay~\cite{Nahrstedt09,Stoica11}. In this subsection, we define a utility model only for the quality of the video signal while during the analysis of our optimization framework we discuss our approach for minimizing the effects of delay. The main objective of our video quality model is to capture the rate-distortion (RD) relationship of different representations of each video stream. This will allow our optimization framework to allocate resources to videos depending on their quality.

In this paper we assume we have the RD information information for each frame $n$ that belongs to representation $r$ of video $i$ and consists of its size $S_{irn}$ in bits and the importance of the frame for the overall reconstruction quality of the video denoted as $q_{irn}$~\cite{ChakareskiAWTG:04c}. In practice, $q_{irn}$ is the total decrease in the mean square error (MSE) distortion that will affect the video if the frame is decoded by the video player~\cite{chou:packetized-media}. The value of the MSE in $q_{irn}$ includes both the distortion that is added when frame $n$ is not decoded, and also the frames that have a decoding dependency with $n$.\footnote{For example $q$ for an I frame includes the $q$ of the P and B frames that depend on it.} Hence, the video quality model considers also the possible drift that might occur due to the inability to decode a particular video frame. These values can be obtained easily but only during the offline encoding of the video as discussed in~\cite{freris13,video-traces}.

Consequently, the aggregate video quality of a group of video frames indexed by $s$ (also referred to as segment to ensure consistency with DASH terminology), that belong to representation $r$ of video $i$, is the average \text{MSE reduction/frame}:
\begin{align}\label{eqn:utility_segment}
 q_{irs}=   \frac{\sum_{n} q_{irn}}{\text{number of frames} } 
\end{align}
This fraction is the average MSE reduction of the frames contained in a DASH segment or packet, versus their total number. \emph{This formulation is in line with our initial objective since it expresses the "value" for a group of frames.} For a group of segments starting from $t$-th segment until the end of the video, we can similarly characterize the video quality as:
\begin{align}\label{eqn:utility_sequence}
    Q_{irt}= \frac{\sum^{s=\text{last}}_{s=t} q_{irs}}{\text{number of remaining segments}} 
\end{align}
For packet-based video, this RD information associated with a packet can be contained in each packet header. In the case of scalable video the information about the importance of a packet is already embedded in the header since it indicates the video layer that the packet belongs. For segment-based DASH streaming a media presentation description (MPD) file is already used for conveying a subset of this information~\cite{dash-mpd}. Hence, the model can support packetized non-scalable, scalable, and segment-based video. The final result of the previous discussion is that a single video for user $i$ will be available at the following discrete set of qualities
\begin{equation}\label{eq:utility-set}
\mathcal{Q}_{it}=\{Q_{i1t},..,Q_{irt},..., \}~\text{with }r\in~\mathcal{R}_{i}
\end{equation}
indicating the set of available representations for each user/video $i$. It is important to understand the use of the previous model in our optimization. In our initial framework, where the problem is solved for the complete playback duration of the video, the formulation in~\eqref{eqn:utility_sequence} is used by setting $t$=0, i.e., we use the average quality of the complete video. However, for DASH the optimization is solved during a specific time period $t$ and ~\eqref{eqn:utility_sequence} captures the video quality of the remaining segments that still need to be communicated.

To complete our discussion, we have to recall that our optimization targets a heterogeneous user population where a subset of them do not receive video. When we have elastic flows, or when the users do not participate in the video-aware optimization, then rate allocation is exercised with a PFRA metric~\cite{andrews14b}, i.e., $Q_{irt}$ is generated by taking the logarithm of the communication rate achievable by user $i$.

\subsection{Throughput at the Physical Layer}
\label{section:throughput-estimation}
We consider that the BSs optimize independently the PHY parameters of the point-to-point links, as it is typically done in wireless communication systems~\cite{book:fundamental-wireless}. To estimate the average communication rate that each user $i$ achieves when it is associated to BS $j$ we proceed as follows. The BS receives from each user $i$ the average channel gain $\Exp [|h_{j,i}|^2]$, and also the local estimate of the interference power $I_{\text{ABS},i}$, $I_{\text{RS},i}$.\footnote{LTE Rel. 8 already implements the communication of the power of the local interference though the high interference indicator (HII).} Note that the average channel gains, or the CQI in the LTE terminology, mentioned above can be transmitted from each user to the BS with low network overhead since they only correspond to path loss and shadowing. Hence, during an ABS, since a user receives the aggregate interference from all the simultaneously transmitting PBSs, the average SINR between the PBS and user $i$ is:
\begin{equation}
\label{ineq:1}
\Exp  [\gamma^\text{ABS}_i]=\frac{P_\text{PBS}\Exp  [|h_{\text{PBS},i}|^2]}{I_{\text{ABS},i}+\sigma^2}
\end{equation}
During the RS the MBS is also active and so the SINR of users associated to a PBS and the MBS are
\begin{equation*}
                   \Exp [ \gamma^{\text{RS}}_{i} ]=\frac{P_\text{PBS}\Exp [ |h_{\text{PBS},i}|^2] }{I_{\text{RS},i}+\sigma^2},~\text{and}~\Exp [ \gamma^{\text{RS}}_{i} ]=\frac{P_\text{MBS} \Exp [ |h_{\text{MBS},i}|^2 ] }{I_{\text{RS},i}+\sigma^2},
\end{equation*}
respectively. The average SINR expressions allow each BS $j$ to estimate the resulting average data rate for each associated user $i$ under MCS $m$ as:
\begin{equation}
C_{im}=m \cdot eff \cdot S \cdot (1-P_s)^{L/m} \text{ bits/sec},
\end{equation}
where $S$ is the symbol rate, $eff$ is the efficiency of the MCS, and the probability of symbol error $P_s$ under $2^m$-QAM is~\cite{proakis}:
\begin{align}
P_s=4(1-2^{m/2})Q \Big ( \sqrt{\frac{3}{2^{m}-1}\Exp [\gamma_{i}]} \Big )
\end{align}
In our system the PHY and link-layer system at each BS selects optimally for the average SINR the MCS that ensures the highest point to point communication rate~\cite{argyriou-tmm08}. This is formally written as:
\begin{equation}
\label{eqn:mcs_selection}
C_{i}=\max_{m\in\mathcal{M}} C_{im}
\end{equation}

\section{Problem Formulation and Solution}
\label{section:optimization}
Now we are ready to define formally the problem we address in this paper. For each user $i$ associated to BS $j$ the HCN must select the video representation with the highest quality, and the rate allocated to it. For the complete HCN the globally optimal TDRP must be calculated. First we define the optimization variables. Let $x^{\text{ABS}}_{ir},x^{\text{RS}}_{ir}\in\{0,1\}$ indicate whether user $i\in \mathcal{N}_j\cup\mathcal{F}_j$ is served with video representation $r$ in an ABS and RS respectively. Let also $z^{\text{ABS}}_{ir}\in [0,1]$ denote the fraction of the ABS resources that the PBS allocates to $i\in\mathcal{N}_j\cup\mathcal{F}_j$ for streaming the video representation $r$. Similarly for the RS, we define $z^{\text{RS}}_{ir}\in [0,1]$. Hence, the decisions of each BS $j$ are: (a) the \textit{video quality selection (VQS)} vector for all the associated users, i.e., $\bm{x}_j=\big(x^{\text{ABS}}_{ir}\geq 0:i\in\mathcal{N}_j\cup\mathcal{F}_j,r\in\mathcal{R}_i)$, and (b) the \textit{rate allocation (RA)} vector for all users, i.e., $\bm{z}_j^{\text{ABS}}=\big(z^{\text{ABS}}_{ir}\geq 0:i\in\mathcal{N}_j\cup\mathcal{F}_j,r\in\mathcal{R}_i)$. Similarly the VQS and RA vectors for the regular slots. Also the global resource partitioning decision $\eta$. To minimize the notation later in our solution, we also define different concatenations of the variable vectors as follows: $\bm{z}_j=\big(\bm{z}_j^{\text{ABS}},\bm{z}_j^{\text{RS}})$, $\bm{z}=\big(\bm{z}_{j}:j\in\mathcal{J})$, and similarly for $\bm{x}_j,\bm{x}$.

The objective for the HCN operator is to maximize the average aggregate delivered video quality captured by:

 \vspace{-0.3cm}
 \small
 \begin{equation}
 \label{eq:BS-objective}
    \sum_{j\in\mathcal{J} \backslash \{0\}}\sum_{i\in\mathcal{N}_j\cup\mathcal{F}_j}\sum_{r\in\mathcal{R}_i}  ( x^{\text{ABS}}_{ir}   + x^{\text{RS}}_{ir})Q_{ir0}+ \sum_{i\in\mathcal{N}_0\cup\mathcal{F}_0}\sum_{r\in\mathcal{R}_i} x^{\text{RS}}_{ir} Q_{ir0}
 \end{equation}
 \normalsize
In the above recall that $Q_{ir0}$ is the average quality of representation $r$ for video $i$. Thus, the objective expresses the video quality delivered to the complete HCN. In the second term we have the quality for the users associated to the MBS since they cannot transmit during an ABS.

For the first set of constraints we have to recall that the fraction of the blank ABS resources available for the PBSs (there is resource re-use across the PBSs) is $\eta$. This leads to:
\begin{equation}
\sum_{ i\in\mathcal{N}_j\cup\mathcal{F}_j}\sum_{r\in\mathcal{R}_i}z^{\text{ABS}}_{ir}\leq \eta,\forall j\in\mathcal{J}\backslash\{0\} \label{eq:zConstraint}
\end{equation}
In the above we excluded again the MBS since it cannot transmit during an ABS. During the RS all the BSs transmit:
\begin{equation}
\sum_{ i\in\mathcal{N}_j\cup\mathcal{F}_j}\sum_{r\in\mathcal{R}_i}z^{\text{RS}}_{ir} \leq 1-\eta, \forall j\in\mathcal{J} \label{eq:zConstraint2}
\end{equation}
When a particular representation $r$ is selected, the average rate $R_{ir}$ in bits/sec that must be sustained by a user $i$ is less than the rate that can be achieved during both the ABS and RS. Also the resources allocated during ABS and RS will determine the average rate. The above can be formally written as:
\begin{equation}
\label{eq:zx1Constraint}
x^{\text{ABS}}_{ir} R_{ir}\leq ( z^{\text{ABS}}_{ir}C^{\text{ABS}}_{i} +z^{\text{RS}}_{ir} C^{\text{RS}}_{i}),\forall r\in\mathcal{R}_i,i\in\mathcal{N}_j\cup\mathcal{F}_j,j\in\mathcal{J}
\end{equation}
\normalsize
In this section this constraint is based in~\eqref{eqn:Rir}, and in this form it ensures that the average number of rebuffering time over the complete video playback is zero. Also,~\eqref{eq:zx1Constraint} accounts for the startup delay as specified in~\eqref{eqn:Rir}. We will delve into the extension for DASH in the next section.

We also have that resources cannot be allocated to a video representation $r$ if it is not actually selected:
\begin{align}
& z^{\text{ABS}}_{ir} \leq x^{\text{ABS}}_{ir},\forall r\in\mathcal{R}_i,i\in\mathcal{N}_j\cup\mathcal{F}_j,j\in\mathcal{J}
\label{eq:zxConstraint}\\
& z^{\text{RS}}_{ir} \leq x^{\text{RS}}_{ir},\forall r\in\mathcal{R}_i,\forall i\in\mathcal{N}_j\cup\mathcal{F}_j,j\in\mathcal{J}
\label{eq:zxConstraint2}
\end{align}
We also need the integer constraints according to which only one video representation $r$ can be used for each user. Thus:
\begin{align}
& \sum_{r\in\mathcal{R}_{i}}x^{\text{ABS}}_{ir} \leq 1, \forall i\in \mathcal{N}_j\cup\mathcal{F}_j,j\in\mathcal{J}
\label{eq:x2Constraint}\\
& \sum_{r\in\mathcal{R}_{i}}x^{\text{RS}}_{ir} \leq 1, \forall i\in \mathcal{N}_j\cup\mathcal{F}_j,j\in\mathcal{J}
\label{eq:x2Constraint2}
\end{align}
During the regular slots the PBSs can also transmit together with the MBS, albeit with lower spectral efficiency. In this case the rate will be lower. However, we must ensure that across ABS and RS the same video representation is used:
\begin{equation}
x^{\text{ABS}}_{ir}=x^{\text{RS}}_{ir},\forall r\in\mathcal{R}_i,\forall i\in \mathcal{N}_j\cup\mathcal{F}_j,j\in\mathcal{J}
\label{eq:xxConstraint}
\end{equation}
The last condition comes from the observation that it is not practical to transmit one video quality during ABS and a different during the RS, since these two types of resources alternate at the PHY in the order of milliseconds~\cite{LTE12}.

\subsection{Hierarchical Primal and Dual Decomposition}
This problem formulation clearly constitutes a non-convex mixed integer linear program (MILP). Hence, it is NP-hard while it does not map to a well-known structure that can be solved with fast pseudo-polynomial algorithms (e.g., knapsack forms). This last aspect can also be seen fairly easily since we have that $\eta$, which is the capacity of the knapsack in~\eqref{eq:zConstraint},\eqref{eq:zConstraint2}, is also an optimization variable. The second issue is the evident need for distributed computation. These aspects make the problem computationally challenging. Despite this difficult challenge at first sight, we notice that at this stage of our work we are interested to calculate the average rate allocation and TDRP during the complete streaming session. This means that in practice the final algorithm does not have to provide a solution in the order of seconds, but minutes. For this reason, instead of designing heuristics, we resort to a solution approach with a primal-dual approximation algorithm that converges asymptotically to the optimal solution~\cite{bertsekas-convex}.

To solve this problem we apply first primal decomposition on $\eta$. Primal decomposition consists of setting a constant value to the coupling variable~\cite{decomposition-tutorial}. For a constant value for $\eta$ we notice that the original problem is decomposed into a master problem $\text{P}_0$, and several problems denoted as $\text{P}_j$ (each one for each BS $j$). We use Lagrangian relaxation to solve $\text{P}_j$. The Lagrangian after relaxing the coupling constraints for $\text{P}_j$ is:

 \begin{align}
L_j&=     \sum_{i\in\mathcal{N}_j\cup\mathcal{F}_j}\sum_{r\in\mathcal{R}_i} (x^{\text{ABS}}_{ir} + x^{\text{RS}}_{ir} )Q_{ir0}
   + \bm{\lambda}_{1,j}^{\text{ABS}} \bm{f}^{\text{ABS}}_1(\bm{z}^{\text{ABS}}_j) \nonumber\\
   &+\bm{\lambda}_{1,j}^{\text{RS}} \bm{f}^{\text{RS}}_1(\bm{z}^{\text{RS}}_j)+\bm{\eta} (\bm{\lambda}_{1,j}^{\text{RS}}-\bm{\lambda}_{1,j}^{\text{ABS}})-\bm{\lambda}_{1,j}^{\text{RS}}\bm{1}
+\bm{\lambda}_{2,j} \bm{f}_2(\bm{z}^{\text{ABS}}_{j}) \nonumber\\
&+\bm{\lambda}_{2,j} \bm{f}_2(\bm{z}^{\text{RS}}_{j})
   +\bm{\lambda}_{2,j} \bm{f}_2(\bm{x}^{\text{ABS}}_{j})
+\bm{\lambda}_{3,j}^{\text{ABS}} \bm{f}^{\text{ABS}}_3(\bm{z}^{\text{ABS}}_j)\nonumber\\
    &+\bm{\lambda}_{3,j}^{\text{ABS}} \bm{f}^{\text{ABS}}_3(\bm{x}^{\text{ABS}}_{j}
+    \bm{\lambda}_{3,j}^{\text{RS}} \bm{f}^{\text{RS}}_3(\bm{z}^{\text{RS}}_j)
    +\bm{\lambda}_{3,j}^{\text{RS}} \bm{f}^{\text{RS}}_3(\bm{x}^{\text{RS}}_{j})\nonumber\\
    &+\bm{\lambda}_{4,j}^{\text{ABS}} \bm{f}^{\text{ABS}}_4(\bm{x}^{\text{ABS}}_j)+\bm{\lambda}_{4,j}^{\text{RS}} \bm{f}^{\text{RS}}_4(\bm{x}^{\text{RS}}_j)\nonumber\\
    &+\bm{\lambda}_{5,j} \bm{f}_5(\bm{x}^{\text{ABS}}_{j})+\bm{\lambda}_{5,j} \bm{f}_5(\bm{x}^{\text{RS}}_j)
    \label{eqn:lagrangian}
    \end{align}
    \normalsize
In the above $\bm{f}^{\text{ABS}}_1,\bm{f}^{\text{RS}}_1$ are constraint vectors~\eqref{eq:zConstraint},~\eqref{eq:zConstraint2}, constraint~\eqref{eq:zx1Constraint} is expressed with vector $\bm{f}_2$, and~\eqref{eq:zxConstraint},\eqref{eq:zxConstraint2} are written as the constraint vectors~$\bm{f}^{\text{ABS}}_3$,$\bm{f}^{\text{RS}}_3$, while $\bm{f}^{\text{ABS}}_4$,$\bm{f}^{\text{RS}}_4$ correspond to~\eqref{eq:x2Constraint},\eqref{eq:x2Constraint2}. Finally $\bm{f}_5$ corresponds to~\eqref{eq:xxConstraint}. The dual variables $\bm{\lambda}$ are all in row vector form in order to avoid the need for a transpose superscript. Now by using this relaxation, and by packing all the Lagrangian multipliers for PBS $j$ as the single vector $\bm{\lambda}_j$, the dual problem can be written as follows:
\begin{equation}
\min_{\bm{\lambda}_j}\max_{\bm{z}_j,\bm{x}_j} L_j(\bm{z}_j,\bm{x}_j,\bm{\lambda}_j)~
\text{s.t.}~\eqref{eq:zConstraint}-\eqref{eq:xxConstraint},\bm{\lambda}_j\geq 0 \label{eq:primal-dual}
\end{equation}
A constant $\eta$ after primal decomposition has further implications. In particular the problem in \eqref{eq:primal-dual} is decomposed into subproblems for the ABS and RS. We also notice from $L_j$ that these subproblems can be further decomposed for both the ABS and the RS. First we have a RA problem $\text{P}^\text{ABS-RA}_j$:
\begin{align}
\label{eq:primal-ra}
 & \max_{\bm{z}^{\text{ABS}}_j}  ( \bm{\lambda}_{1,j}^\text{ABS} \bm{f}^\text{ABS}_1(\bm{z}^{\text{ABS}}_j)-\bm{\eta} \bm{\lambda_{1,j}}^{\text{ABS}} +\bm{\lambda}_{2,j} \bm{f}_2(\bm{z}^{\text{ABS}}_j) \nonumber \\
&+\bm{\lambda}_{3,j}^\text{ABS} \bm{f}^\text{ABS}_3(\bm{z}^{\text{ABS}}_j))~\text{s.t.}~\eqref{eq:zConstraint}, \eqref{eq:zx1Constraint},\eqref{eq:zxConstraint}
\end{align}
\normalsize
Also a VQS problem, denoted as $\text{P}^\text{ABS-VQS}_j$:
\begin{align}
\label{eq:primal-vqtms}
&\max_{\bm{x}^\text{ABS}_j} (\sum_{i\in\mathcal{N}_j\cup\mathcal{F}_j} \sum_{r\in\mathcal{R}_i} x^{\text{ABS}}_{ir}Q_{ir0}+\bm{\lambda}_{2,j} \bm{f}_2(\bm{x}^{\text{ABS}}_{j})+ \bm{\lambda}_{3,j} \bm{f_3}(\bm{x}^{\text{ABS}}_{j})\nonumber\\
&+\bm{\lambda}_{4,j} \bm{f_4}(\bm{x}^{\text{ABS}}_j)+\bm{\lambda}_{5,j} \bm{f_5}(\bm{x}^{\text{ABS}}_j) )~\text{s.t.} ~\eqref{eq:zx1Constraint},\eqref{eq:zxConstraint},\eqref{eq:x2Constraint},\eqref{eq:xxConstraint}
\end{align}
\normalsize
Thus, we have two linear RA and two integer VQS problems that are solved by each BS as we explain next.

\subsection{Rate Allocation and Video Quality Selection at the BSs}
The dual problem is solved in an iterative fashion, using a primal-dual Lagrange method that can allow us to reach an asymptotically optimal solution~\cite{bertsekas-convex,poularakis-infocom}. The central concept of the primal-dual algorithm is to initialize first the dual variables to zero, and then to solve subproblems $\text{P}^\text{ABS-RA}_j,\text{P}^\text{ABS-VQS}_j,\text{P}^\text{RS-RA}_j,\text{P}^\text{RS-VQS}_j$ to obtain the currently optimal solution for iteration $\tau$ as $\bm{z}_j(\tau)$, and $\bm{x}_j(\tau)$ . Besides our problem-specific details described in this subsection, further details regarding the primal-dual method and its asymptotically optimal convergence property can be found in~\cite{bertsekas-convex,poularakis-infocom}.

First we focus on subproblem~\eqref{eq:primal-ra} that is linear program. Hence, it can be efficiently solved using standard convex optimization techniques~\cite{bertsekas-convex}. The BS also solves the second subproblem, i.e., the integer linear program (ILP) in~\eqref{eq:primal-vqtms} for identifying the currently optimal video representation for each user. This problem is solved in pseudo-polynomial time using dynamic programming (DP)~\cite{bertsekas-convex}. Its speed of convergence is evaluated collectively for the complete system in our performance evaluation, while the time complexity of the complete JTRAVQS is discussed in a later subsection.

After solving the subproblems, and given the current result $\bm{z}^{\text{ABS}}_j(\tau),\bm{x}^{\text{ABS}}_j(\tau)$, we employ a sub-gradient method \cite{bertsekas-convex} to update the dual variables. A representative calculation is presented for the dual variables of the very first constraint:
\begin{equation}
\lambda_{1,jir}(\tau+1)=\Big [\lambda_{1,jir}(\tau)+\beta(\tau)  (  z^{\text{ABS}}_{ir}(\tau)
-\eta  ) \Big ]^+
\label{eq:dual_update2}
\end{equation}
In the above, the term in the parenthesis is the sub-gradient, $[.]^+$ denotes the projection onto the non-negative orthant, and $\beta(\tau)$ is the step size at iteration $\tau$. Similarly we define the update rules and the subgradients for the remaining dual variables. In each iteration $\tau$, the dual objective is improved using the subgradient update and accordingly the primal relaxed problems $\text{P}^\text{ABS-RA}_j,\text{P}^\text{ABS-VQS}_j,\text{P}^\text{RS-RA}_j,\text{P}^\text{RS-VQS}_j$ are solved again in order to update the primal variables (which are then used in the subsequent dual objective update).

\subsection{Solving the Master Problem}
For the final step, the dual variables for constraints of each $\text{P}_j$ are provided to the master problem $\text{P}_0$ that has to be solved now for the optimal $\eta$ at the central controller (CC) of the system. Recall that for the master problem we applied primal decomposition. It is thus solved very efficiently by collecting only the resource prices for $\eta$, i.e $\lambda^{\text{ABS}}_{1,j}(\tau),\lambda^{\text{RS}}_{1,j}(\tau)$, from each PBS in order to form the global subgradient~\cite{decomposition-tutorial}. In practice we only transmit the local subgradient $\lambda^{\text{ABS}}_{1,j}(\tau)-\lambda^{\text{RS}}_{1,j}(\tau)$ from each BS. The current value of $\eta$ is updated as follows~\cite{decomposition-tutorial}:

\vspace{-0.3cm}
\small
\begin{equation}
\eta(\tau+1)=\Big [ \eta(\tau)+\bm{\beta}(\tau) \underbrace{\left [ \begin{array}{ccc}
   0-\lambda^{\text{RS}}_{1,0}(\tau)\\
   ...\\
  \lambda^{\text{ABS}}_{1,j}(\tau)-\lambda^{\text{RS}}_{1,j}(\tau)\\
  ...
  \end{array}\right]   }_\text{global subgradient}\Big ]^+
\label{eq:primal_update}
\end{equation}
\normalsize
Now $\bm{\beta}(\tau)$ is a vector that can be selected as before in order to control the speed of convergence~\cite{bertsekas-convex}.

\subsection{Discussion on Time Complexity}
The time complexity of the discrete knapsack problem denoted by $\text{P}^\text{ABS-VQS}_j$, when solved with dynamic programming, is polynomial with respect to the size of the problem $|\mathcal{N}_j||\mathcal{R}|$ for each BS $j$, but for bounded knapsack capacity. Hence, in our case it is $\mathcal{O}(|\mathcal{N}_j|\cdot|\mathcal{R}|\cdot 1)$, i.e., linear. This is because from~\eqref{eq:x2Constraint} we notice that the capacity of the knapsack is 1, in other words only one item (video representation) can be selected. $\text{P}^\text{ABS-RA}_j$  is a LP and so polynomial with respect to the number of associated users $|\mathcal{N}_j|$, i.e., $\text{P}^\text{ABS-RA}_j$ is $\mathcal{P}(|\mathcal{N}_j|)$. Hence, the time complexity of  $\text{P}^\text{ABS-VQS}_j$ may be better than that of $\text{P}^\text{ABS-RA}_j$. We conclude that the worst execution time of one iteration of the JTRAVQS algorithm, as a function of its inputs, can be expressed as:\footnote{The quantities in the $\mathcal{O},\mathcal{P}$ notation have to be multipled with the execution time of the fundamental algorithm operation.}

\vspace{-0.3cm}
\small
\begin{equation*}
\max_{j \in \mathcal{J}} \Big ( \max \big ( \mathcal{O}(|\mathcal{N}_j||\mathcal{R}|),\mathcal{P}(|\mathcal{N}_j|) \big )+d_{j,CC} \Big )+\mathcal{O}(1)+\max_{j \in \mathcal{J}} ( d_{CC,j} ),
\end{equation*}
\normalsize
where $d_{j,CC}$ is the delay between BS $j$ and the CC, $\mathcal{O}(1)$ corresponds to the execution time of the primal update (one vector multiplication and one addition) and $d_{CC,j}$ is the delay for communicating the new primal update to the BSs. In our simulation we also considered that the backhaul links have the \emph{worst case} transmission delay of $d_{j,CC}$=60ms~\cite{backhaul-smallcell}. For 100 iterations the total delay until the optimal solution is reached is 6 seconds which means that the calculations have to take place within 4 seconds in order to reach a solution within a 10 second period. This is well within the capabilities of modern processors for solving the discussed LP and ILP. Also, each BS $j$ communicates only $\lambda^{\text{ABS}}_{1,j},\lambda^{\text{RS}}_{1,j}$ to the CC which constitutes a negligible overhead. The good performance of the algorithm, allow us to investigate its use in shorter time scales next.

\section{Problem Formulation and Solution for DASH}
\label{section:optimization-dash}
\textbf{Motivation.} In a network it is possible that channel conditions and users are more dynamic. In this case the bitrate of the transmitted video should be adapted. One way to accomplish that is DASH. With DASH a video is stored as a sequence of short duration (typically 2-10 sec) video segments~\cite{dash}. Each segment may be available at different sizes, SNR qualities, spatial resolutions, frame rates. However, it has been shown that allowing the client to be fully responsible for requesting video segments (a pull-based system), after estimating the variations in the end-to-end throughput, results in significant waste of resources~\cite{Schulzrinne13,Ammar14}. In this paper, our optimization framework at the BS is responsible for the choice of the optimal video representation. This is also a realistic option since DASH does not specify where video adaptation occurs.

\begin{figure}[t]%
\centering
\includegraphics[keepaspectratio,width = 0.95\linewidth]{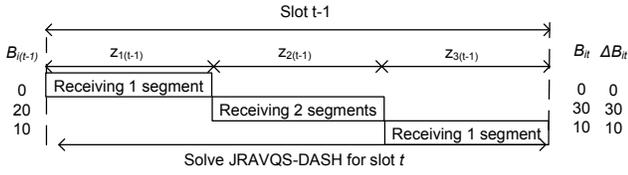}
\caption{Transmission of different segments during slot $t-1$ that has a durationof 10 seconds. JTRAVQS-DASH is solved to reach the decisions for the next slot $t$.}
\label{fig:enforcement-of-ra}
\end{figure}

\textbf{Enhanced System Model.} We define the term \textit{slot} as the period that the problem is solved and its decisions are enforced (see Fig.~\ref{fig:enforcement-of-ra}). Without loosing generality we assume that JTRAVQS-DASH is solved during a slot with a duration of 10 seconds. Since the problem is solved for every slot, the instance of the problem currently solved is also indexed by $t$. The result is that the JTRAVQS-DASH problem is solved during slot $t$ to calculate the optimal RA and VQS for slot $t+1$. The algorithm requires several iterations as before, that are indexed also by $\tau$. Regarding the input parameters to JTRAVQS-DASH $C_{it}$ is our estimate of the TCP throughput of user $i$ during slot $t$ according to~\cite{argyriou-pv07}, and $\mathcal{N}_{jt}$ the set of associated users. This approach for modeling $C_{it}$, is consistent with the behavior of DASH that uses TCP for downloading segments. Hence, contrary to related work our approach is more realistic with respect to $C_{it}$~\cite{veciana14}. The video quality model in~\eqref{eqn:utility_sequence} is used with $Q_{irt}$ denoting the quality of the remaining segments that should be transmitted. Let us finally define some minimal additional notation since the optimization variables must be indexed by slot $t$: $\bm{z}_{jt}=\big(z^{\text{ABS}}_{irt},z^{\text{RS}}_{irt}\geq 0:i\in\mathcal{N}_{jt}\cup\mathcal{F}_{jt},r\in\mathcal{R}_i)$ and $\bm{x}_{jt}=\big(x^{\text{ABS}}_{irt}\geq 0:i\in\mathcal{N}_{jt}\cup\mathcal{F}_{jt},r\in\mathcal{R}_i)$.  Now $x^{\text{ABS}}_{irt}$ indicates that in slot $t$ segments from representation $r$ will be transmitted to user $i$. The same algorithm is used for solving JTRAVQS-DASH but the two subproblems are adapted.

\textbf{DASH Rate Allocation (DASHRA) Problem.} The most important aspect is the re-formulation of constraint~\eqref{eq:zx1Constraint} that is now indexed by the slot $t$, and is packed into constraint vector $\bm{f}^\text{DASH}_2$:
\small
\begin{equation}
\label{eq:zx1Constraint_DASH}
x^{\text{ABS}}_{irt} S_{irt}\leq ( z^{\text{ABS}}_{irt}C^{\text{ABS}}_{it} +z^{\text{RS}}_{irt} C^{\text{RS}}_{it}) \max \{ \Delta B_{it}, 1\},
\forall r\in\mathcal{R}_i,i\in \mathcal{N}_{jt}
\end{equation}
\normalsize
Re-writing \eqref{eq:primal-ra} for the DASH case yields the problem $\text{P}^\text{ABS-DASHRA}_{jt}$:
\begin{align}%
\label{eqn:RART_opt}
 & \max_{\bm{z}^{\text{ABS}}_{jt}} (\bm{\lambda}_{1,j}^\text{ABS} \bm{f}^\text{ABS}_1(\bm{z}^{\text{ABS}}_{jt})
-\bm{\eta} \bm{\lambda}^\text{ABS}_{1,j}+\bm{\lambda}_{2,j} \bm{f}^\text{DASH}_2(\bm{z}^{\text{ABS}}_{jt},\bm{\Delta B}_{t})\nonumber\\ &+\bm{\lambda}_{3,j}^\text{ABS} \bm{f}^\text{ABS}_3(\bm{z}^{\text{ABS}}_{jt}))~\text{s.t.}~\eqref{eq:zConstraint}, \eqref{eq:zx1Constraint_DASH},\eqref{eq:zxConstraint}
\end{align}
\normalsize
In~\eqref{eq:zx1Constraint_DASH} $S_{irt}$ is the average bitrate that must be sustained by the remaining segments of the $r$-th representation similarly to~\eqref{eqn:Rir},~\eqref{eqn:utility_sequence}. The key difference from the initial problem is parameter $\Delta B_{it}$. This is the total duration of the playable video in seconds that user $i$ has in its buffer at the start of slot $t$ and is denoted as $B_{it}$, minus the playable video that the slowest user has in its buffer, i.e., $\Delta B_{it}= B_{it}-B_{(\text{slowest})t}$. Each user updates the estimate of the playable video as: $B_{it}=B_{i(t-1)}+\text{received during ($t$-1)}-\text{played during ($t$-1)}$. This parameter ensures that users that have received lower volume of data are effectively prioritized. Hence, if the RA decision for user $i$ in the $t-1$-th slot is $z_{i(t-1)}$, then at the start of slot $t-1$ the BS can calculate $B_{it}$, since it knows the result of RA and of course the duration of video that will be played. To summarize, this is effectively a rebuffering constraint that contains the differential buffer information.

\textbf{DASH Client Model Example.} To explain the playback model for the DASH client and the setting of $\Delta B_{it}$, let us use a specific example with clients that have different playback buffer contents as illustrated in Fig.~\ref{fig:enforcement-of-ra}. The number of transmitted segments depend on the value of $z_{i(t-1)}$. Also in this example we consider the downloading of complete segments for exposition purposes but our model supports partially downloaded segments. For user 1 assume that it is the client that is lagging behind from the rest and the playble video it has in its buffer at the start of slot $t-1$ is $B_{1(t-1)}$=0. Hence, at the start of slot $t-1$ it will rebuffer until it receives the segment and after it finishes, the video player enters the playback mode. Also $B_{1t}$=$B_{1(t-1)}$+10-10=0, and $\Delta B_{1t}$=$B_{1t}-B_{1t}$=0. At the start of slot $t-1$ user 2 has 20 seconds worth of video, while during $t-1$ it will receive two additional segments leading to $B_{2t}$=20+20-10=30, and $\Delta B_{2t}$=$B_{2t}-B_{1t}$=30. Hence, by having allocated more resources with $z_{2(t-1)}$ the result is a pre-fetching of data. For user 3 similarly we obtain $B_{3t}$=10+10-10=10 and $\Delta B_{2t}=B_{3t}-B_{1t}$=10.

\textbf{DASH Video Quality Selection (DASHVQS) Problem.} Now the VQS problem is solved by adding one constraint in the problem~\eqref{eq:primal-vqtms}. We reformulate the $\text{P}^\text{ABS-VQS}_{j}$  problem to $\text{P}^\text{ABS-DASHVQS}_{jt}$  that is also solved over the $t-1$ slot:

\small
\begin{align*}%
\label{eqn:VQSRT_opt}
& \max_{\bm{x}^\text{ABS}_{jt}}  (\sum_{i\in\mathcal{N}_j\cup\mathcal{F}_j} \sum_{r\in\mathcal{R}_i} x^{\text{ABS}}_{irt}Q_{irt}+\bm{\lambda}_{2,j} \bm{f}^\text{DASH}_2(\bm{x}^{\text{ABS}}_{jt})+ \bm{\lambda}_{3,j} \bm{f_3}(\bm{x}^{\text{ABS}}_{jt})\\
&+\bm{\lambda}_{4,j} \bm{f_4}(\bm{x}^{\text{ABS}}_{jt})+\bm{\lambda}_{5,j} \bm{f_5}(\bm{x}^{\text{ABS}}_{j})  )
~~\text{s.t.} ~\eqref{eq:zx1Constraint},\eqref{eq:zxConstraint},\eqref{eq:x2Constraint},\eqref{eq:xxConstraint}
\end{align*} %
\normalsize
Note that for partial downloading, if in the next slot DASHVQS identifies that a lower or higher video quality is transmitted, then pre-buffered data are not discarded but the unfinished segment is received. The new decision for the video quality is enforced when a new segment will be transmitted. The dual variables are updated as before and the sub-gradients are similarly modified based on the new constraint.

\section{Performance Evaluation}
\label{section:performance-evaluation}
In this section, we present a comprehensive evaluation of the proposed algorithms comprising our framework through custom Matlab simulation. Our simulator performs a precise PHY-level simulation of wireless packet transmissions.

\textbf{JTRAVQS Evaluation.} The parameter settings for our simulations are set as follows. Downlink MBS and PBS transmit power are equal to 46dBm and 30dBm respectively~\cite{andrews14b}. Distance-dependent path loss is given by $L(d)=128.1 + 37.6\log_{10}(d)$, where $d$ is the distance between two nodes in Km~\cite{LTE12}, and the shadowing standard deviation is 8 dB. The user speed is 3 kmph (quasi-static as we already stated), and average CQI is provided every 10 minutes. The macrocell area is set to be a circle with radius equal to 1 Km. The wireless channel parameters include a channel bandwidth of $W$=20 MHz, noise power spectral density of $\sigma^2$=$10^{-6}$ Watt/Hz, while the same Rayleigh fading model was used for all the channels. Packets of 1500 bytes are transmitted at the PHY, while the optimal MCS is calculated according to~\eqref{eqn:mcs_selection}. The user distribution and picocell locations are random and uniform within the macrocell.  We set the biasing threshold to 0 dB for all the systems to calculate $\mathcal{N}_j$. The user population increases up to a number of 200 to evaluate the performance in networks that continuously become more dense, consistently with the recent trends~\cite{andrews14b}. For the PFRA system we configured the users to request randomly and uniformly one of the available video representations, while for JTRAVQS users request randomly and uniformly one of the available videos. The video content used in the experiments consists of 26 CIF (352x288), and high definition (1920x1080) sequences that were encoded with SVC H.264 as a single layers~\cite{video-traces}. The videos are compressed at 30 fps and different rates ranging from 128 Kbps and reaching values$<$7 Mbps. The frame-type patterns were G16B1,G16B3,G16B7,G16B15, i.e., there are different numbers of B frames between every two P frames and a GOP size is always equal to 16 frames.
\begin{figure}[t]
\centering
\subfigure[]
{\includegraphics[keepaspectratio,width = 0.495\linewidth]{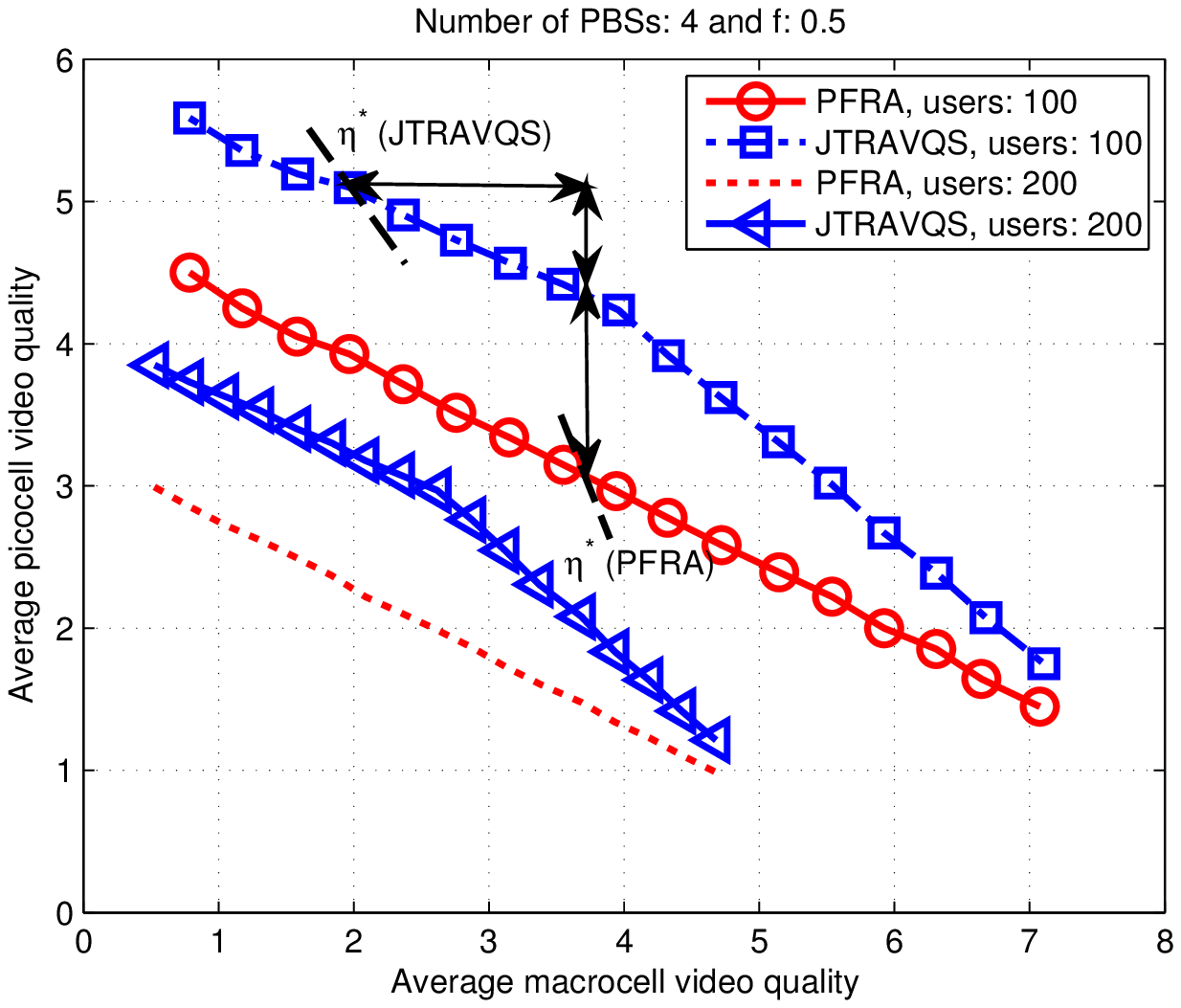}}\hspace{-0.2cm}%
\subfigure[]{\includegraphics[keepaspectratio,width = 0.495\linewidth]{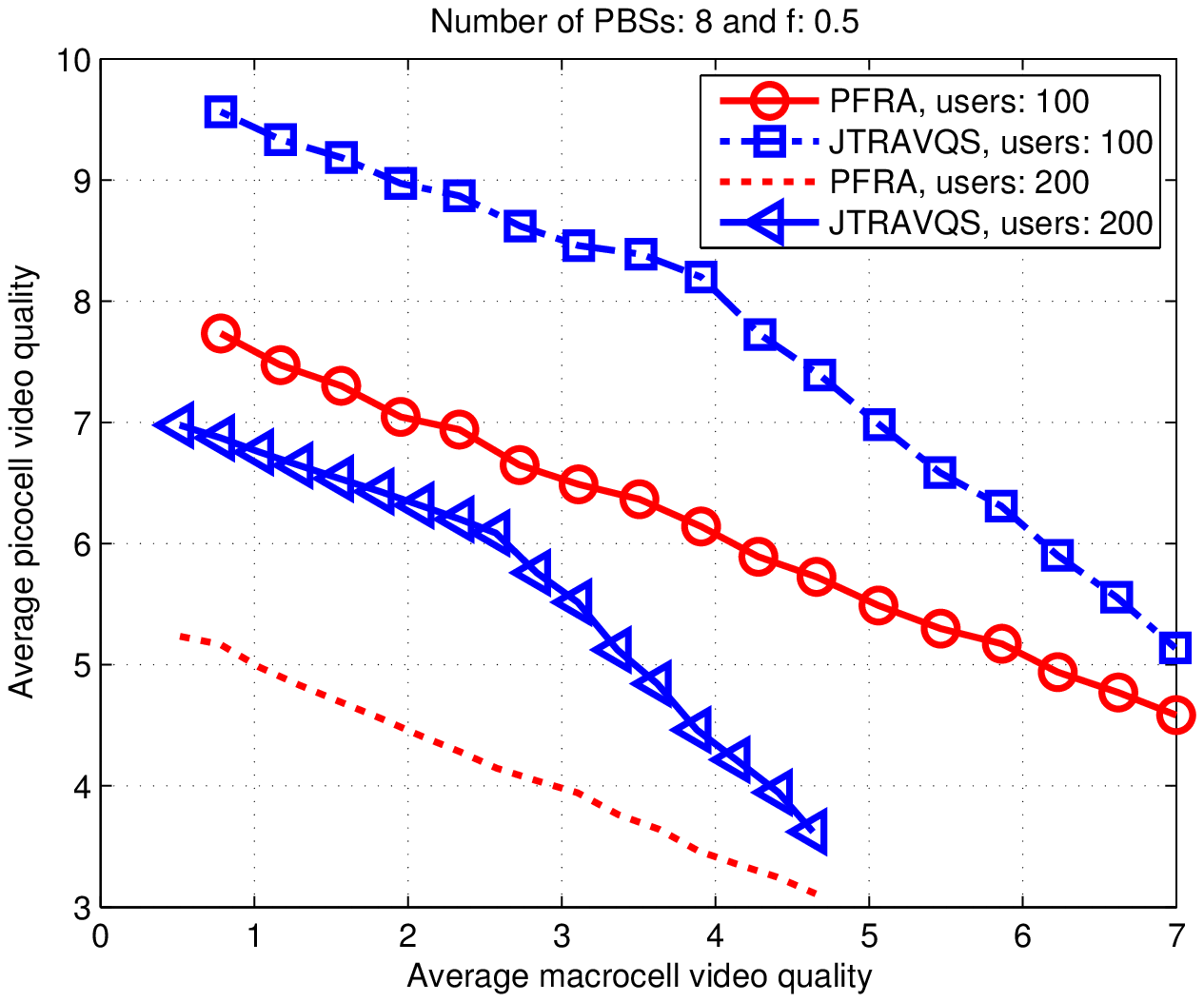}}\hspace{-0.2cm}%
\subfigure[]{\includegraphics[keepaspectratio,width = 0.495\linewidth]{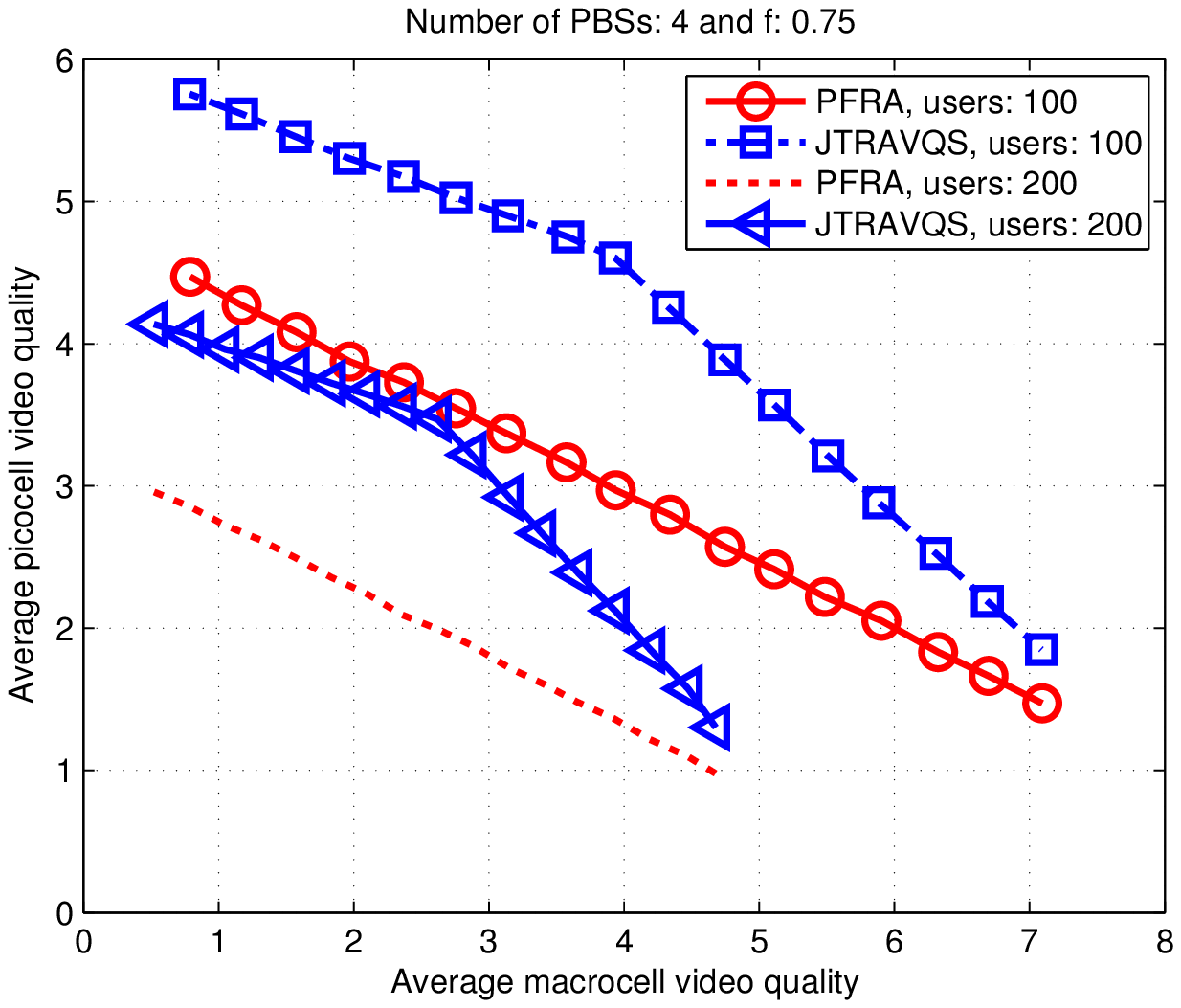}}\hspace{-0.2cm}%
\subfigure[]{\includegraphics[keepaspectratio,width = 0.495\linewidth]{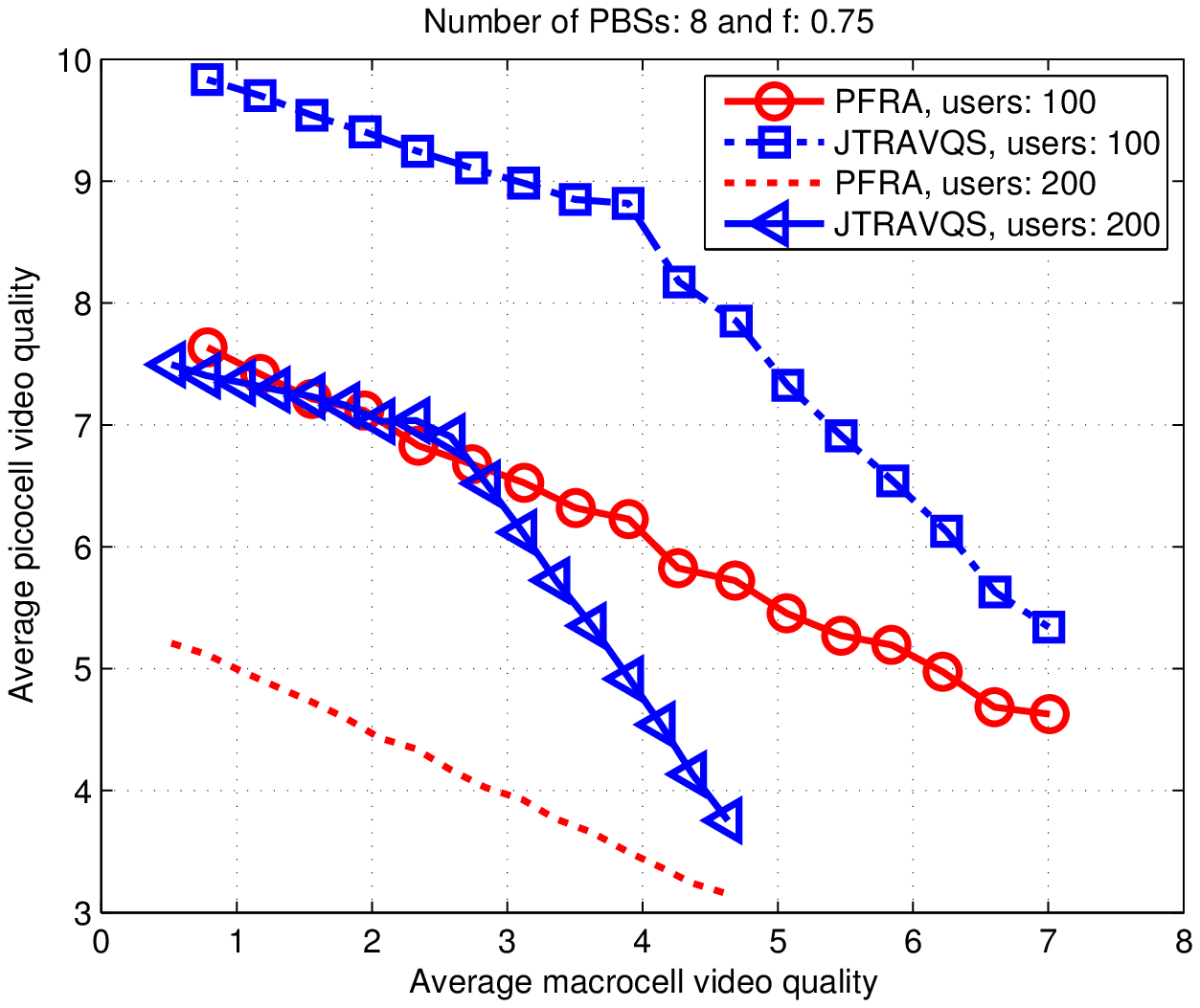}}%
\caption{Average macrocell vs. aggregate picocell video quality. }
\label{fig:results2}
\end{figure}

Regarding the presentation of the results, Fig.~\ref{fig:results2} shows the average video quality (in terms of the representation $r$) that is delivered to the picocell and macrocell users. For example one data point that has the value 3.2 indicates that on average the users received the quality representation 3.2. Hence, higher values indicate that the users received on average higher video quality representations. The data points in these figures correspond to different values of $\eta$. Also, the data points correspond to the average (mean) of all the measurements for 100 randomly generated topologies. The sample variance for this set the measurements is between 0.1 and 0.2 which is fairly small compared to the value of the mean and its difference between all the tested systems.

\textbf{Video Quality.} For this set of results we present the average video quality of the picocell users versus the average video quality of the users associated to the macrocell (only from the MBS to its associated users) for different constant values of $\eta$ to illustrate the impact of different TDRP. The results for all systems can be seen in Fig.~\ref{fig:results2}(a,b) for $f$=0.5. JTRAVQS is superior when compared to PFRA for high user density and low PBS density in Fig.~\ref{fig:results2}(a). As the number of the PBSs is increased to 8 in Fig.~\ref{fig:results2}(b), all the systems can achieve higher performance. The reason is that fewer users are associated to each picocell and so a higher communication rate is available for each user under any scheme. So more picocells leads to better results due to the higher available rate per user as expected. Another important result is that for constant PBS density (either 4 or 8), we have higher gain as the user population grows. The reason is the higher importance of optimal rate allocation, since the rate of a single PBS is shared among several users. For example in the very left data point of Fig.~\ref{fig:results2}(b) performance improvement of JTRAVQS over PFRA is 21\% for 100 users and 36\% for 200 users). Of course the average video quality is reduced for all systems since more users are present. Also note that in the left part of the $x$ axis, where all the resources are practically allocated to the picocells ($\eta\approx 1$), we observe the maximum possible video quality in the network. In this regime, the performance gap between JTRAVQS and the other systems is increased as the number of picocells and users is increased.

Another important result in the same figure, is related to the optimal $\eta^*$. It is indicated with a dashed line that is intersected with representative performance curves. This shows that the interpretation of the optimal TDRP with JTRAVQS, that is denoted as $\eta^*$(JTRAVQS), results in higher value for $\eta^*$ when compared to $\eta^*$(PFRA) by 22\% (highlighted with the horizontal arrow). Also if we assume that the system executes first PFRA to calculate the optimal TDRP indicated as $\eta^*$(PFRA), and then perform RAVQS~\cite{argyriou-cnf-hetnet-video} with this fixed value, the result is an average quality equal to 4.3 (the gain is highlighted with the lower vertical arrow). However, our complete system calculates the optimal operating point indicated with $\eta^*$(JTRAVQS) in the figure which gives an average quality equal to 5.1, a performance difference of  18.6\% (the gain is highlighted with the upper vertical arrow).

Our system offers significant performance increase for $f$=0.5 but the benefits are more important when $f$=0.75 in Fig.~\ref{fig:results2}(c,d). Note that for $f$=0.75 the slope of the curve is reduced as the $\eta$ is decreased. The benefit is because we have a higher number of users that can be optimized under JTRAVQS. Also in this case the benefits are even more important when the fraction of the resources $1-\eta$ that the macrocell uses is below 50\% (left part of the $x$ axis) since this gives more resources to the highly spectral efficient links in the picocells to be used and so a higher communication rate is possible.

\begin{figure}[t]
\centering
\subfigure[]{\includegraphics[keepaspectratio,width = 0.499\linewidth]{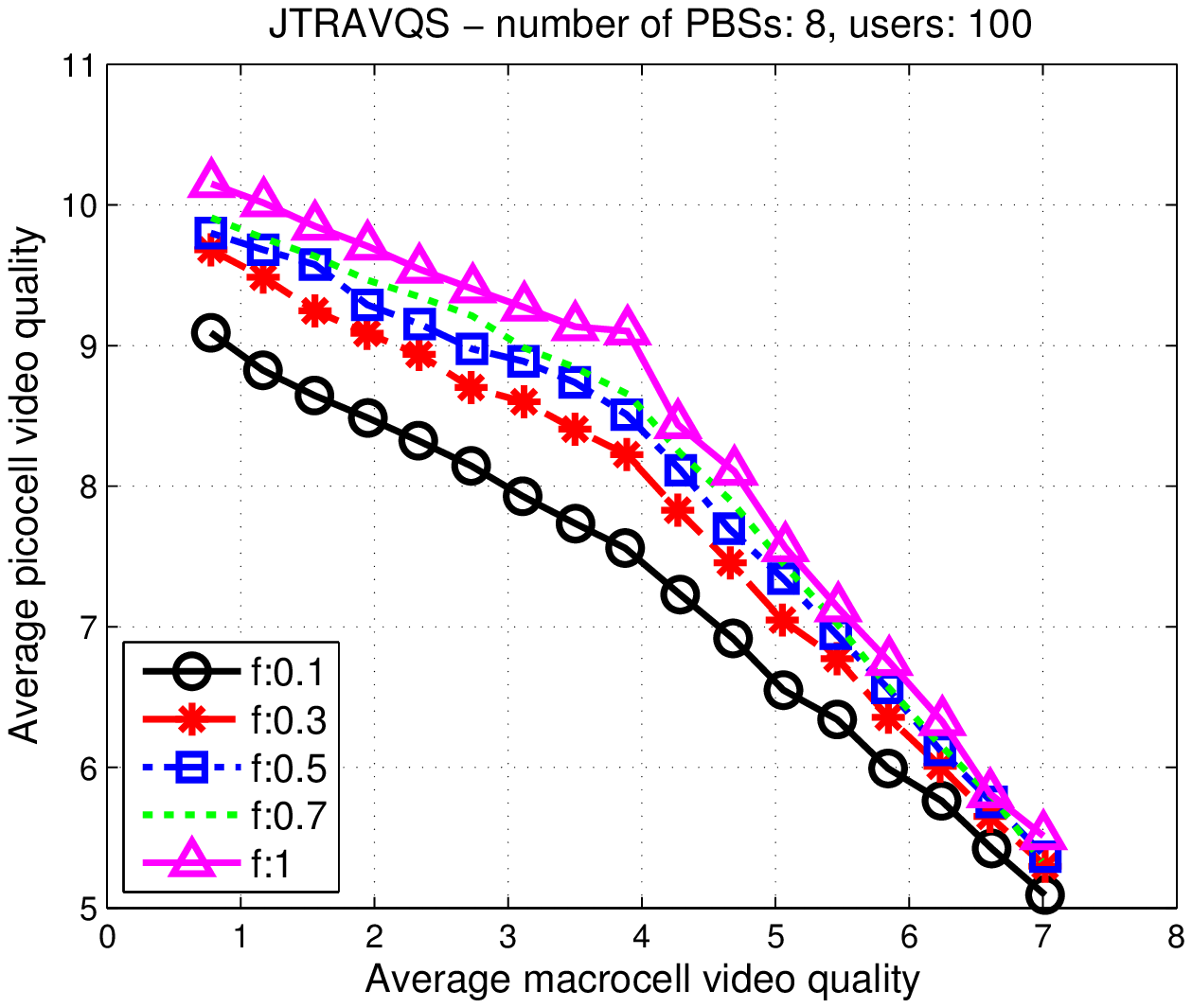}}\hspace{-0.2cm}%
\subfigure[]{\includegraphics[keepaspectratio,width =0.499\linewidth]{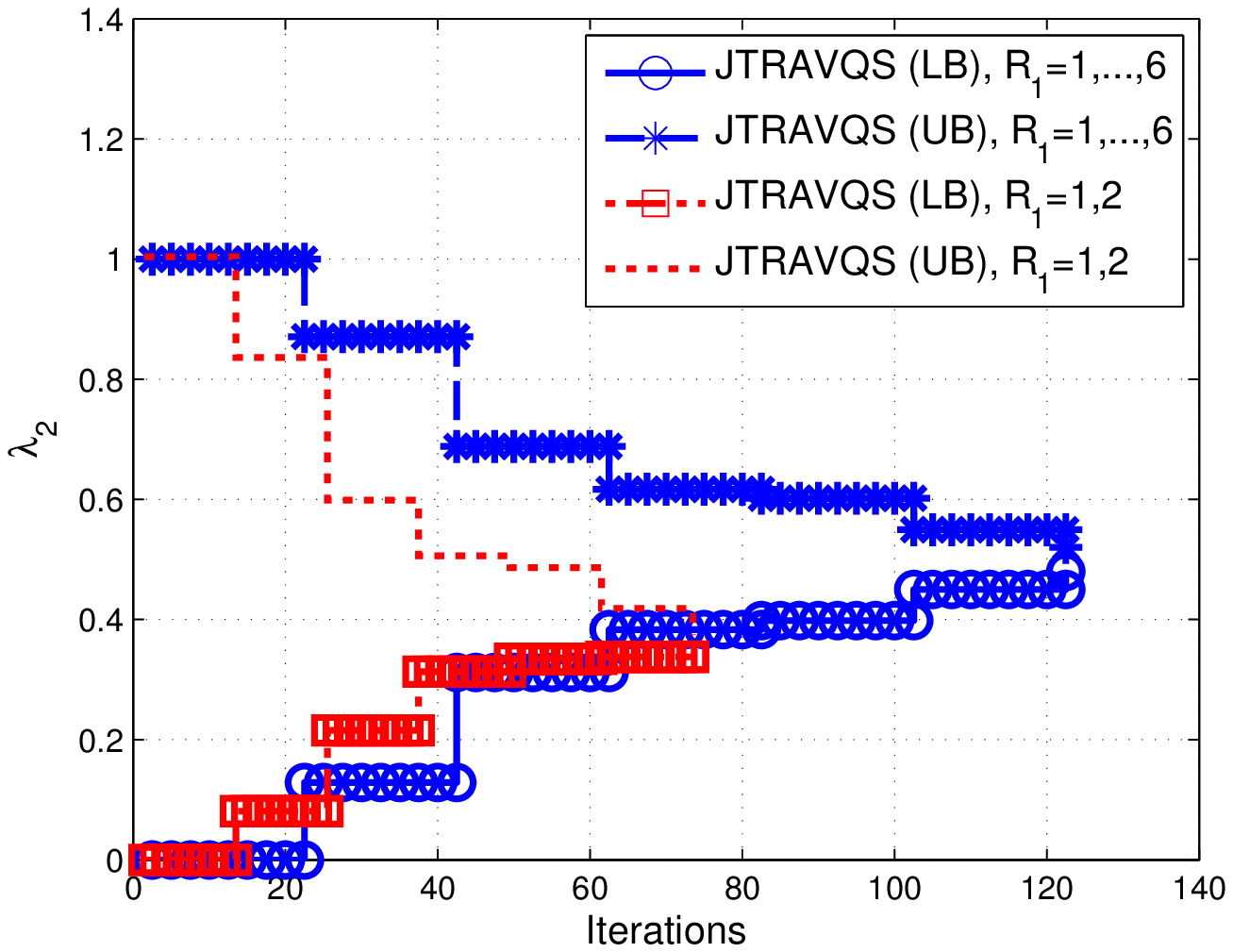}}\hspace{-0.2cm}%
\subfigure[]{\includegraphics[keepaspectratio,width = 0.49\linewidth]{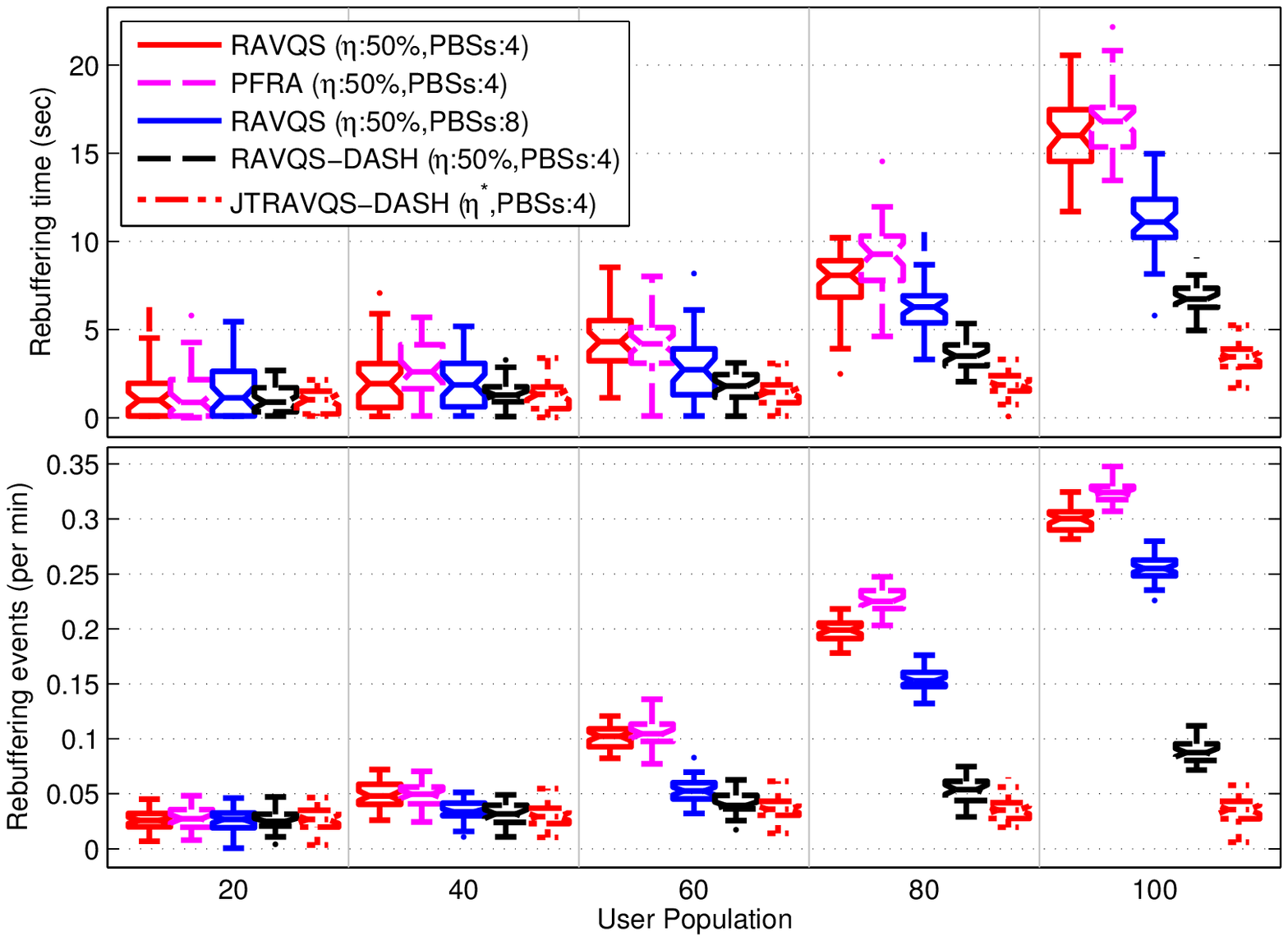}}\hspace{-0.2cm}%
\subfigure[]{\includegraphics[keepaspectratio,width = 0.49\linewidth]{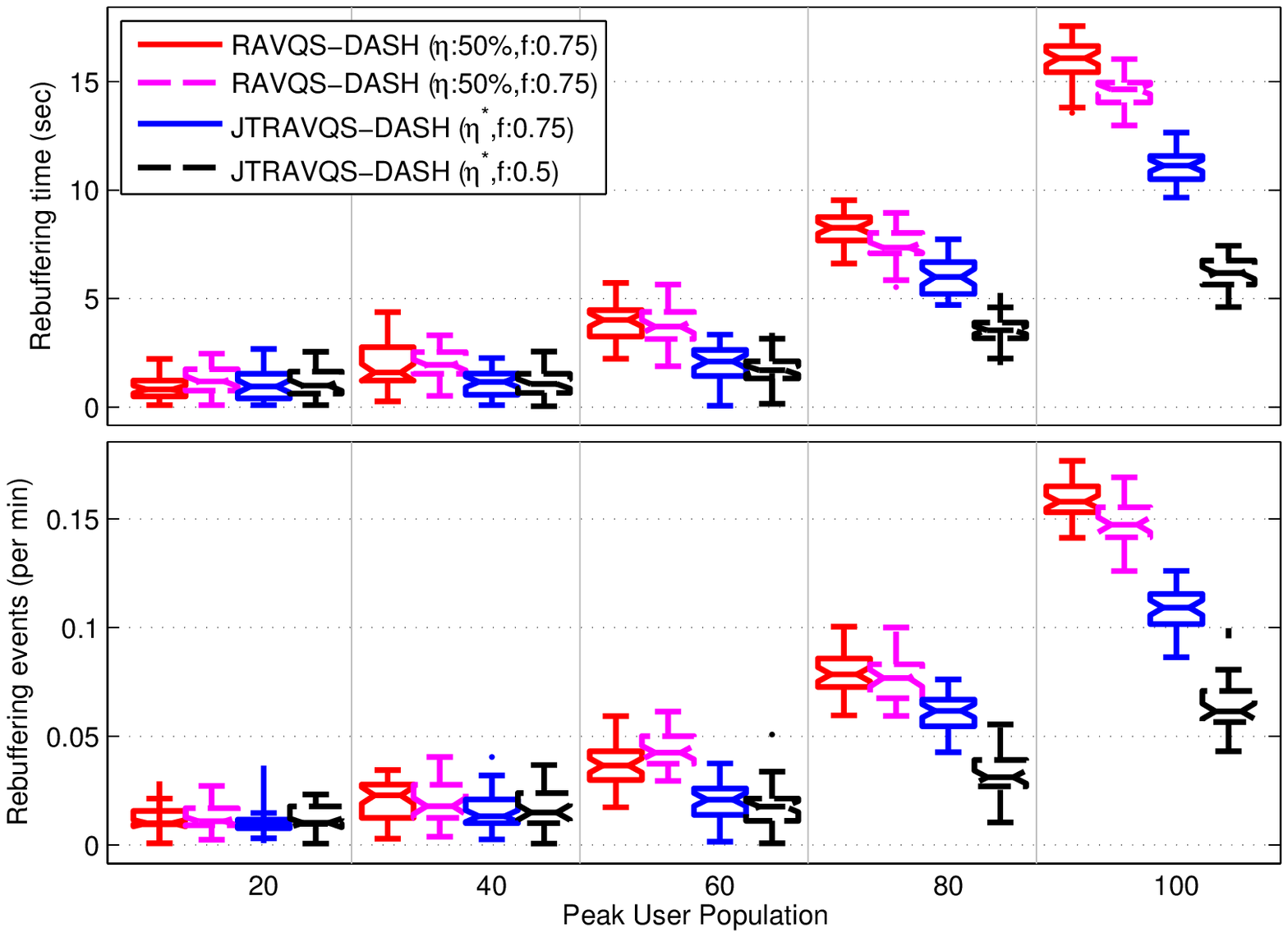}}%
\caption{(a) Video quality for different fraction of participating users. (b) Micro-benchmark for the upper bound (UB) and lower bound (LB) versus the number of iterations for JTRAVQS and a specific user. Two different sets of available video quality representations are shown ($|\mathcal{R}_1|$=2 and $|\mathcal{R}_1|$=6). (c) Rebuffering time/events vs. total number of users. (d) Rebuffering time/events vs. peak number of users with user churn.}
\label{fig:results6}
\end{figure}

\textbf{Fraction of Optimized Users.} Now an interesting set of results is obtained for different values of $f$. We notice in Fig.~\ref{fig:results6}(a) that as this fraction is reduced, JTRAVQS essentially degenerates to the PFRA system. Nevertheless, we still obtain significant benefits even for ratios of $f$ around 30\% since few users are enough for JTRAVQS to be able to improve the overall system performance.

\textbf{Primal-Dual Convergence.} The convergence speed of JTRAVQS versus the number of iterations is illustrated in Fig.~\ref{fig:results6}(b). Results for the algorithm execution are shown for a specific fixed number of picocells and user population. The results for the primal-dual algorithm used for JTRAVQS show that the convergence is achieved within 150 iterations. Also Fig.~\ref{fig:results6}(b) illustrates another important aspect of our system: If the system uses fewer discrete quality representations for each video file, it allows the faster convergence of the algorithm.

\textbf{JTRAVQS-DASH Evaluation.} The performance of JTRAVQS-DASH is evaluated with the same setup as before, that is however augmented when necessary. Now we present results for the playback performance: The time that a client is rebuffering in seconds, and the number of rebuffering events per minute. To ensure fairness, we calculate first the optimal solution with JTRAVQS. Then, the minimum video quality for the JTRAVQS-DASH system is set equal to JTRAVQS. This ensures that JTRAVQS-DASH delivers at least the same video quality and the question is then to evaluate its ability to minimize rebuffering.\footnote{One can plot a synthetic metric of the two but this is not easily interpreted in terms of real QoE.}

For static user conditions the results are illustrated in Fig.~\ref{fig:results6}(c). We draw a notched box plot of the rebuffering time for all the clients in the upper part of Fig.~\ref{fig:results6}(c), and the number of rebuffering events in the lower part of Fig.~\ref{fig:results6}(c). The notch here marks the 95\% confidence interval for the \emph{median}.\footnote{Note that we plot the median here instead of the mean to avoid being influenced by outliers.} All the systems perform well for 20 and 40 users since high capacity is available in the network. However, for higher user densities the buffering time with the baseline JTRAVQS is increased by a factor that is worse than linear. The same is true for PFRA that is slightly worse. With a higher number of 8 PBSs, rebuffering time is improved for JTRAVQS because of higher available capacity. The same is true for the number of rebuffering events/minute that is a very high number for the first three systems we discussed (2-3 events in a 10 minute period). Hence, the higher capacity in the network achieved with 8 PBSs, simply delays the inevitable sharp increase but only of the rebuffering time. This result has an interesting interpretation for MNOs: With increasing user density, expanding the network with more small cells improves marginally the rebuffering time and practically not at all the rebuffering events. This is in contrast to the video quality that can achieve significant improvement in our earlier plots. For extra gains, solutions with buffer-awareness are required.

Better results are obtained for RAVQS-DASH again with a constant $\eta$=50\%. This is effectively a system configuration that encompasses the main features of the DASH-aware streaming literature for single cell networks, e.g.~\cite{veciana14}, where the rate allocation and video quality are optimized by considering the buffer contents, but the overall communication resources are constant. Recall than $\eta$=50\% is the optimal $\eta$ under a PFRA metric for a population of 100 users and 4 PBSs (as illustrated in our earlier figures). Buffer-awareness can indeed reduce the rebuffering when compared to the previous systems, while it can also reduce the variations of playback buffering for the users (we have more predictable performance). The proposed JTRAVQS-DASH system illustrates that it can reduce the time spent in rebuffering by over 50\% when compared to the previous system, while the number of rebuffering events is reduced even more significantly (1 event/25 min. vs. 1 event/10 min.). Hence, for a HCN the fixed TDRP is not the best option even if we design a DASH-aware system. Also, our overall results indicate that using a fixed TDRP has worse consequences in the rebuffering time/events of DASH than on the video quality (e.g., the results illustrated in Fig.~\ref{fig:results2}).

Finally, we evaluate our system for a time-varying user population based on results from a real 3G network reported in~\cite{woo13}. We simulate an 8 hour period between 4pm and 12am, with a user peak occurring around 9pm~\cite{woo13}. During this peak, the number of users is nearly 30\% higher than the number of users at 4pm and 12am. Hence, in our simulation we set accordingly $|\mathcal{N}_{jt}|$ (the increase and decrease are approximated as linear in time as shown in~\cite{woo13}). In the results in Fig.~\ref{fig:results6}(d) we present in the $x$ axis the peak number of users. $C_{it}$ for each user $i$ is also affected since TCP shares equally the communication rate among the competing traffic. Also, we only compare different flavors of JTRAVQS-DASH since the previous systems are not designed for a dynamic network. First, we observe again that the systems with fixed resource partitioning $\eta$=50\% have worse performance. Second, we notice  that the rebuffering time is higher when the fraction of optimized users is also high and equal to $f$=0.75. This means that with increasing user density in a network with user churn, optimizing a lower fraction $f$ of the users increases the DASH playback quality of the optimized users by a significant amount for JTRAVQS-DASH. This is an important result since it provides a tool for an MNO to differentiate QoE in terms of rebuffering to various users.

\section{Conclusions}
\label{section:conclusions}
In this paper, we presented a framework for improving the quality of video streaming in a HCN that employs TDRP. TDRP is essential for the efficient operation of HCNs and when high quality video distribution enters the game, efficiency becomes even more important. Our framework addressed precisely this challenge, i.e., it ensures optimal and video-aware allocation of resources in HCNs that apply TDRP. We formulated this problem in a linear non-convex formulation for which we proposed a primal-dual approximation algorithm. Our problem was decomposed into several problems that included a convex rate allocation problem, and a binary ILP for optimal video quality selection. An extensive performance evaluation under different HCN configurations highlighted the value of our framework for obtaining video quality improvements. Another implication of our solution approach is that it can be solved very fast. This allowed us to augment it to support the more challenging case of DASH. In this case significant additional improvement in terms playback performance was obtained.


\end{document}